\documentclass[superscriptaddress]{revtex4-2}

\usepackage[version=0.96]{pgf}
\usepackage{tikz, calc}
\usetikzlibrary{arrows,shapes,snakes,automata,backgrounds,petri,positioning,fit}
\usepackage{graphicx}

\usetikzlibrary{lindenmayersystems}
\usetikzlibrary[shadings]

\usepackage{float}%float images
\usepackage[font=small,labelfont=bf,
   justification=Justified,
   format=plain]{caption}
\usepackage{subcaption}

\usepackage{amsmath}
\usepackage{amssymb}
\usepackage{hyperref}

\usepackage{xcolor}

\graphicspath{ {images/} }

\def\beq{\begin{equation}}
\def\eeq{\end{equation}}

\def\n{\nu}
\def\be{\begin{equation}}
\def\ee{\end{equation}}
\def\ba{\begin{eqnarray}}
\def\ea{\end{eqnarray}}

\usepackage{graphicx}% Include figure files
\usepackage{dcolumn}% Align table columns on decimal point
\usepackage{bm}% bold math

\begin{document}

\preprint{APS/123-QED}

\title{Efficient Simulation of the 2D Hubbard Model via Hilbert Space-Filling Curve Mapping}

\author{Ashkan Abedi}
\affiliation{Scuola Normale Superiore, NEST, and Istituto Nanoscienze-CNR, 56126 Pisa, Italy}
\author{Vittorio Giovannetti}
\affiliation{Scuola Normale Superiore, NEST, and Istituto Nanoscienze-CNR, 56126 Pisa, Italy}
\author{Dario De Santis}
\affiliation{Scuola Normale Superiore, I-56126 Pisa, Italy}

\begin{abstract}
We investigate tensor network simulations of the two-dimensional Hubbard model by mapping the lattice onto a one-dimensional chain using space-filling curves. In particular, we focus on the Hilbert curve, whose locality-preserving structure minimizes the range of effective interactions in the mapped model. This enables a more compact matrix product state (MPS) representation compared to conventional snake mapping. Through systematic benchmarks, we show that the Hilbert curve consistently yields lower ground-state energies at fixed bond dimension, with the advantage increasing for larger system sizes and in physically relevant interaction regimes. Our implementation reaches clusters up to $32\times32$ sites with open and periodic boundary conditions, delivering reliable ground-state energies and correlation functions in agreement with established results, but at significantly reduced computational cost. These findings establish space-filling curve mappings, particularly the Hilbert curve, as a powerful tool for extending tensor-network studies of strongly correlated two-dimensional quantum systems beyond the limits accessible with standard approaches.
\end{abstract}

\maketitle

\section{\label{sec:level1} Introduction}

The Hubbard model stands as one of the most fundamental paradigms in condensed matter physics, providing an essential framework for understanding strongly correlated electronic systems. Originally proposed by Hubbard in 1963~\cite{hubbard1963}, this deceptively simple model captures the essential physics of electrons moving on a lattice with local Coulomb repulsion, yet exhibits an extraordinarily rich phase diagram encompassing metal-insulator transitions, magnetism, and potentially high-temperature superconductivity~\cite{arovas2022,qin2022}. Despite decades of intensive research, the two-dimensional (2D) Hubbard model continues to resist a complete analytical solution, making numerical approaches indispensable for understanding its physics.

The computational challenge of the 2D Hubbard model derives from the exponential growth of the Hilbert space with system size, making exact diagonalization feasible only for small clusters. Various numerical methods have been developed to tackle this challenge, each with inherent strengths and limitations. Quantum Monte Carlo (QMC) methods~\cite{Matuttis1996,zhang2024afqmc} provide numerically exact results but suffer from the notorious sign problem away from half-filling and at low temperatures, a phenomenon in which the Monte Carlo weights become negative or complex, leading to exponential growth in statistical errors and rendering calculations intractable. Dynamical Mean Field Theory (DMFT) and its cluster extensions~\cite{georges2023dmft,Biroli_2004} capture local correlation effects accurately, but may miss important spatial fluctuations and long-range order. Variational methods including neural network quantum states~\cite{Medvidovi_2024,Bodendorfer_2025} have shown promise but require careful optimization and may suffer from bias in the variational ansatz.

Tensor network methods, particularly the Density Matrix Renormalization Group (DMRG)~\cite{white1992,schollwock2011}, have revolutionized the study of one-dimensional quantum systems by exploiting the area law of entanglement entropy~\cite{Hastings_2007, Eisert_2010}. This motivates our approach of employing efficient one-dimensional Matrix Product State (MPS) algorithms through a suitable mapping. In particular, by projecting the 2D system onto a 1D chain via space-filling curves, the nearest-neighbor interactions are converted into structured long-range couplings that can still be effectively treated within 1D tensor network algorithms~\cite{cataldi2021}.

The challenge with naive 2D-to-1D mappings is that they can introduce very long-range interactions in the 1D representation. A recent study by Baldelli et al.~\cite{baldelli2025} pointed out that a straightforward snake-like mapping leads to long-range hopping terms that cause an exponential growth in computational cost with lattice width, limiting traditional DMRG to very narrow cylinders - quasi-2D geometries where a 2D lattice of width $W$ and length $L \gg W$ is treated with periodic (cylinder) or open (strip) boundary conditions in the width direction (typically width~{6-8}). This fundamental limitation motivates the search for more locality-preserving mappings that can maintain the efficiency of 1D tensor network methods while tackling truly 2D systems.

Recent advances in tensor network algorithms have further enhanced their applicability to 2D systems. The development of improved optimization schemes~\cite{vanderstraeten2019tangent}, and better handling of symmetries~\cite{schmoll2017symmetry} have pushed the boundaries of accessible system sizes and accuracy. Furthermore, the integration of machine learning techniques with tensor networks~\cite{liu2023ml_tn} has opened new avenues for efficient representation and optimization of quantum states.

The choice of mapping from 2D to 1D is crucial for the efficiency of tensor network simulations. Cataldi et al.~\cite{cataldi2021} demonstrated that the Hilbert curve~\cite{hilbert_b} mapping significantly outperforms conventional approaches for the 2D quantum Ising model. The Hilbert curve, a continuous fractal space-filling curve, possesses unique locality-preserving properties that minimize the average distance between neighboring sites after mapping, resulting in more efficient tensor network representations. Furthermore, a recent study by Bellwood et al.~\cite{bellwood2025} examined a wide range of 2D-to-1D path mappings and confirmed that fractal space-filling curves like the Hilbert curve generally yield faster convergence and higher accuracy than the standard snake path, with the benefit increasing for larger systems. They propose methods to construct optimal paths by tiling smaller fractal segments, further validating the approach we adopt here.

In this work, we extend and optimize the Hilbert curve mapping approach for the fermionic 2D Hubbard model, which presents additional challenges due to the presence of both spin and charge degrees of freedom and the need to properly handle fermionic statistics. To our knowledge, this is the first application of the Hilbert curve mapping to a fermionic system with spin-up and spin-down degrees of freedom, extending beyond the spin models studied in previous works~\cite{cataldi2021}. We perform comprehensive benchmarking against the conventional snake mapping and other recent approaches, demonstrating substantial improvements in both accuracy and computational efficiency. Our key contributions include: (i) systematic analysis of how the Hilbert curve preserves locality in the fermionic Hubbard model, (ii) quantitative comparison of convergence rates between different mappings for systems up to $32 \times 32$ sites, and (iii) demonstration that the Hilbert mapping enables more accurate simulations in parameter regimes where traditional approaches struggle.

The remainder of this paper is organized as follows. In Section \ref{sec:preliminaries}, we describe the space-filling curve mappings and a brief introduction to the MPS formalism and the numerical methods we used in our simulations. Section \ref{sec:results} presents our numerical results, comparing ground-state energies and convergence properties for different system sizes and parameter regimes, studying both half-filled and doped cases with a particular focus on the intermediate coupling regime. Finally, Section \ref{sec:conclusions} summarizes our findings and discusses future directions for applying space-filling curve mappings to other strongly correlated systems.

Our results establish the Hilbert curve mapping as a powerful tool for studying 2D strongly correlated systems, offering a favorable balance between accuracy and computational cost. This approach opens new possibilities for investigating the phase diagram of the Hubbard model and related models in parameter regimes that were previously inaccessible to accurate numerical treatment.
\section{\label{sec:preliminaries} Preliminaries}

\subsection{Space-Filling Curves}

Mapping a two-dimensional lattice onto a one-dimensional chain is a prerequisite for applying highly efficient tensor-network algorithms such as the MPS ansatz. The choice of mapping, however, is not merely a technical detail: it determines how local couplings in the original lattice are translated into interactions in the effective 1D Hamiltonian. Poor mappings can transform short-range 2D couplings into long-range 1D terms, increasing entanglement across MPS bonds and inflating the bond dimension required for accurate simulations. Space-filling curves provide a systematic class of mappings that traverse every site of a discrete grid in a continuous path, with varying degrees of locality preservation.

\subsubsection{Snake Mapping}

The simplest example is the \emph{snake} mapping, which proceeds row by row across the lattice, reversing direction at each row boundary (see Figure~\ref{fig:mappings_8x8}). Starting from the bottom-left corner, the path traverses the first row from left to right, then reverses direction to traverse the second row from right to left, and continues alternating until all sites are visited. This yields a bijection from 2D lattice coordinates to 1D chain positions that can be written explicitly as
\begin{equation}
\mathcal{M}_{\text{snake}}(i,j) = 
\begin{cases}
i \cdot n + j + 1, & \text{if } i \text{ is even}, \\
i \cdot n + (n-j), & \text{if } i \text{ is odd},
\end{cases}
\end{equation}
where $(i,j)$ are row and column indices of an $n\times n$ lattice. For example, $\mathcal{M}_{\text{snake}}(0,0)=1$, $\mathcal{M}_{\text{snake}}(0,n-1)=n$, and $\mathcal{M}_{\text{snake}}(1,n-1)=n+1$, which corresponds to the point where the snake path makes its first turn.

While trivial to implement, the snake mapping performs poorly at preserving locality. Vertical neighbors at row boundaries, separated by a single lattice spacing in 2D, may be mapped to positions differing by up to $n-1$ in the 1D chain. This generates hopping terms with effective range $\mathcal{O}(n)$, which significantly increase the Matrix Product Operator (MPO) bond dimension and induce higher entanglement across MPS bonds. As a result, the computational cost grows rapidly with system width, severely limiting scalability.

\subsubsection{Hilbert Curve Mapping}

In contrast, the Hilbert curve is a self-similar, fractal space-filling curve that offers superior locality preservation. Originally introduced in the mathematical literature as a continuous surjection from the unit interval to the unit square, its discrete form provides an ordering of lattice sites that minimizes the distance between mapped neighbors. Crucially, the Hilbert curve is \emph{recursive}: each higher-order curve is built by appropriately rotating and connecting four copies of the lower-order curve. This recursive structure not only ensures continuity but also endows the mapping with a natural hierarchy, making it well suited for tensor-network applications.

\subsubsection{Recursive Construction of the Hilbert Curve}

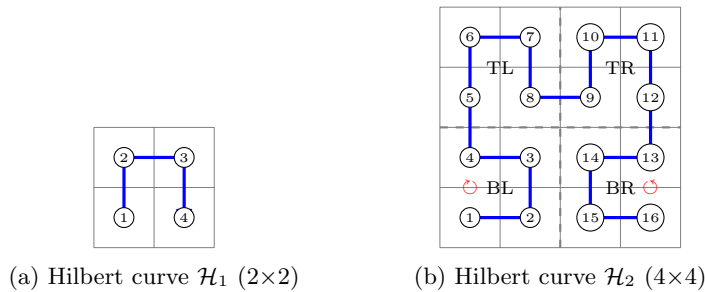
\begin{figure}
\hspace{-1.4cm}
\centering
\begin{tikzpicture}[scale=0.8]
% First order - 2x2
\begin{scope}[xshift=0cm]
\draw[step=1cm,gray,very thin] (0,0) grid (2,2);
\draw[blue,very thick,->] 
(0.5,0.5) -- (0.5,1.5) -- (1.5,1.5) -- (1.5,0.5);
\foreach \x/\y/\n in {0.5/0.5/1, 0.5/1.5/2, 1.5/1.5/3, 1.5/0.5/4} {
    \node[circle,fill=white,draw=black,inner sep=1pt] at (\x,\y) {\tiny \n};
}
\node at (1,-0.5) {(a) Hilbert curve $\mathcal{H}_1$ (2×2)};
\end{scope}
\hspace{1.4cm}
% Second order - 4x4 with quadrant labels
\begin{scope}[xshift=4cm]
\draw[step=1cm,gray,very thin] (0,0) grid (4,4);
\draw[dashed,gray,thick] (2,0) -- (2,4);
\draw[dashed,gray,thick] (0,2) -- (4,2);

% Label quadrants
\node[black] at (1,3.0) {\scriptsize TL};
\node[black] at (3,3.0) {\scriptsize TR};
\node[black] at (1,1.0) {\scriptsize BL};
\node[black] at (3,1.0) {\scriptsize BR};

% Show rotation symbols
\node[red] at (0.5,1) {\scriptsize $\circlearrowright$};
\node[red] at (3.5,1) {\scriptsize $\circlearrowleft$};

\draw[blue,very thick,->] 
(0.5,0.5) -- (1.5,0.5) -- (1.5,1.5) -- (0.5,1.5) --
(0.5,2.5) -- (0.5,3.5) -- (1.5,3.5) -- (1.5,2.5) --
(2.5,2.5) -- (2.5,3.5) -- (3.5,3.5) -- (3.5,2.5) --
(3.5,1.5) -- (2.5,1.5) -- (2.5,0.5) -- (3.5,0.5);
\foreach \x/\y/\n in {0.5/0.5/1, 1.5/0.5/2, 1.5/1.5/3, 0.5/1.5/4,
                      0.5/2.5/5, 0.5/3.5/6, 1.5/3.5/7, 1.5/2.5/8,
                      2.5/2.5/9, 2.5/3.5/10, 3.5/3.5/11, 3.5/2.5/12,
                      3.5/1.5/13, 2.5/1.5/14, 2.5/0.5/15, 3.5/0.5/16} {
    \node[circle,fill=white,draw=black,inner sep=1pt] at (\x,\y) {\tiny \n};
}
\node at (2,-0.5) {(b)  Hilbert curve  $\mathcal{H}_2$ (4×4)};
\end{scope}

\end{tikzpicture}
\caption{Construction of the Hilbert curve. (a) First-order curve $\mathcal{H}_1$ connecting 4 sites. (b) Second-order curve $\mathcal{H}_2$ for a $4 \times 4$ lattice, showing quadrant divisions (TL: top-left, TR: top-right, BL: bottom-left, BR: bottom-right) and rotation directions.}
\label{fig:hilbert_construction}
\end{figure}

The first-order Hilbert curve $\mathcal{H}_1$ connects the four sites of a $2\times 2$ lattice in a U-shaped path (see Figure~\ref{fig:hilbert_construction}(a)). Higher-order curves are obtained by dividing an $n\times n$ lattice ($n=2^k$) into four quadrants, placing rotated copies of $\mathcal{H}_{k-1}$ in each quadrant, and then connecting them. We depict how to construct $\mathcal{H}_2$ as a composition of rotated copies of $\mathcal{H}_1$ in Figure~\ref{fig:hilbert_construction}. The generic description to construct $\mathcal{H}_k$ from $\mathcal{H}_{k-1}$ can be explained through the following three steps:

\begin{enumerate}
    \item \textbf{Partition}: Divide the $n\times n$ lattice into four quadrants of size $n/2 \times n/2$.
    \item \textbf{Orient}: Insert a copy of $\mathcal{H}_{k-1}$ on each quadrant and apply the following rotations:
    \begin{itemize}
        \item Bottom left (BL)  quadrant: $90^\circ$ clockwise rotation of $\mathcal{H}_{k-1}$,
        \item Bottom right (BR) quadrant: $90^\circ$ counterclockwise rotation of $\mathcal{H}_{k-1}$,
        \item Top left (TL) and Top right (TR) quadrants: no rotations applied.
    \end{itemize}
    \item \textbf{Connect}: Join the quadrants sequentially to form a continuous path, entering at the bottom-left corner and exiting at the bottom-right corner of the full lattice.
\end{enumerate}

%%%% figure of hilbert generation

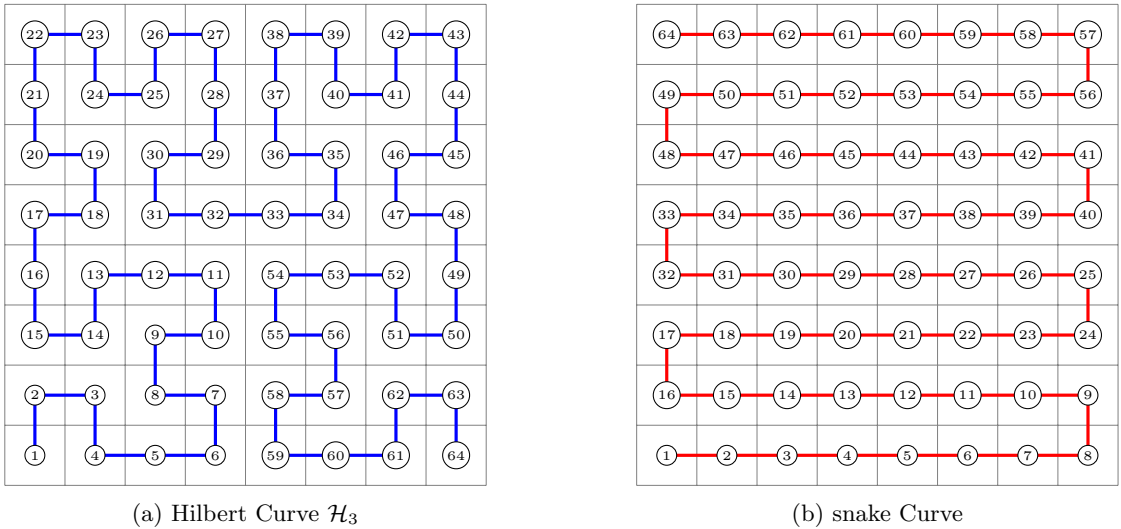
\begin{figure}[htb]
\centering
\begin{tikzpicture}[scale=0.8]

% ------------------ Hilbert curve for 8x8 ------------------
\begin{scope}[xshift=0cm]
  \draw[step=1cm,gray,very thin] (0,0) grid (8,8);

  % Path goes from site 1 to 64 (arrow at 64)
  \draw[blue,very thick,->]
  (0.5,0.5) -- (0.5,1.5) -- (1.5,1.5) -- (1.5,0.5) -- (2.5,0.5) -- (3.5,0.5) -- (3.5,1.5) -- (2.5,1.5) -- (2.5,2.5) -- (3.5,2.5)
  -- (3.5,3.5) -- (2.5,3.5) -- (1.5,3.5) -- (1.5,2.5) -- (0.5,2.5) -- (0.5,3.5) -- (0.5,4.5) -- (1.5,4.5) -- (1.5,5.5) -- (0.5,5.5)
  -- (0.5,6.5) -- (0.5,7.5) -- (1.5,7.5) -- (1.5,6.5) -- (2.5,6.5) -- (2.5,7.5) -- (3.5,7.5) -- (3.5,6.5) -- (3.5,5.5) -- (2.5,5.5)
  -- (2.5,4.5) -- (3.5,4.5) -- (4.5,4.5) -- (5.5,4.5) -- (5.5,5.5) -- (4.5,5.5) -- (4.5,6.5) -- (4.5,7.5) -- (5.5,7.5) -- (5.5,6.5)
  -- (6.5,6.5) -- (6.5,7.5) -- (7.5,7.5) -- (7.5,6.5) -- (7.5,5.5) -- (6.5,5.5) -- (6.5,4.5) -- (7.5,4.5) -- (7.5,3.5) -- (7.5,2.5)
  -- (6.5,2.5) -- (6.5,3.5) -- (5.5,3.5) -- (4.5,3.5) -- (4.5,2.5) -- (5.5,2.5) -- (5.5,1.5) -- (4.5,1.5) -- (4.5,0.5) -- (5.5,0.5)
  -- (6.5,0.5) -- (6.5,1.5) -- (7.5,1.5) -- (7.5,0.5);

  % site numbers 1..64 on centers
  \foreach \x/\y/\n in {
    0.5/0.5/1, 0.5/1.5/2, 1.5/1.5/3, 1.5/0.5/4, 2.5/0.5/5, 3.5/0.5/6, 3.5/1.5/7, 2.5/1.5/8,
    2.5/2.5/9, 3.5/2.5/10, 3.5/3.5/11, 2.5/3.5/12, 1.5/3.5/13, 1.5/2.5/14, 0.5/2.5/15, 0.5/3.5/16,
    0.5/4.5/17, 1.5/4.5/18, 1.5/5.5/19, 0.5/5.5/20, 0.5/6.5/21, 0.5/7.5/22, 1.5/7.5/23, 1.5/6.5/24,
    2.5/6.5/25, 2.5/7.5/26, 3.5/7.5/27, 3.5/6.5/28, 3.5/5.5/29, 2.5/5.5/30, 2.5/4.5/31, 3.5/4.5/32,
    4.5/4.5/33, 5.5/4.5/34, 5.5/5.5/35, 4.5/5.5/36, 4.5/6.5/37, 4.5/7.5/38, 5.5/7.5/39, 5.5/6.5/40,
    6.5/6.5/41, 6.5/7.5/42, 7.5/7.5/43, 7.5/6.5/44, 7.5/5.5/45, 6.5/5.5/46, 6.5/4.5/47, 7.5/4.5/48,
    7.5/3.5/49, 7.5/2.5/50, 6.5/2.5/51, 6.5/3.5/52, 5.5/3.5/53, 4.5/3.5/54, 4.5/2.5/55, 5.5/2.5/56,
    5.5/1.5/57, 4.5/1.5/58, 4.5/0.5/59, 5.5/0.5/60, 6.5/0.5/61, 6.5/1.5/62, 7.5/1.5/63, 7.5/0.5/64
  }{
    \node[circle,fill=white,draw=black,inner sep=1pt] at (\x,\y) {\tiny \n};
  }

  \node at (4,-0.5) {(a) Hilbert Curve $\mathcal{H}_3$};
\end{scope}

% ------------------ snake curve for 8x8 ------------------
\begin{scope}[xshift=10.5cm]
  \draw[step=1cm,gray,very thin] (0,0) grid (8,8);

  % Path goes from site 1 to 64 (arrow at 64)
  \draw[red,very thick,->]
  (0.5,0.5) -- (1.5,0.5) -- (2.5,0.5) -- (3.5,0.5) -- (4.5,0.5) -- (5.5,0.5) -- (6.5,0.5) -- (7.5,0.5) -- (7.5,1.5) -- (6.5,1.5)
  -- (5.5,1.5) -- (4.5,1.5) -- (3.5,1.5) -- (2.5,1.5) -- (1.5,1.5) -- (0.5,1.5) -- (0.5,2.5) -- (1.5,2.5) -- (2.5,2.5) -- (3.5,2.5)
  -- (4.5,2.5) -- (5.5,2.5) -- (6.5,2.5) -- (7.5,2.5) -- (7.5,3.5) -- (6.5,3.5) -- (5.5,3.5) -- (4.5,3.5) -- (3.5,3.5) -- (2.5,3.5)
  -- (1.5,3.5) -- (0.5,3.5) -- (0.5,4.5) -- (1.5,4.5) -- (2.5,4.5) -- (3.5,4.5) -- (4.5,4.5) -- (5.5,4.5) -- (6.5,4.5) -- (7.5,4.5)
  -- (7.5,5.5) -- (6.5,5.5) -- (5.5,5.5) -- (4.5,5.5) -- (3.5,5.5) -- (2.5,5.5) -- (1.5,5.5) -- (0.5,5.5) -- (0.5,6.5) -- (1.5,6.5)
  -- (2.5,6.5) -- (3.5,6.5) -- (4.5,6.5) -- (5.5,6.5) -- (6.5,6.5) -- (7.5,6.5) -- (7.5,7.5) -- (6.5,7.5) -- (5.5,7.5) -- (4.5,7.5)
  -- (3.5,7.5) -- (2.5,7.5) -- (1.5,7.5) -- (0.5,7.5);

  % site numbers 1..64 on centers (snake ordering)
  \foreach \x/\y/\n in {
    0.5/0.5/1, 1.5/0.5/2, 2.5/0.5/3, 3.5/0.5/4, 4.5/0.5/5, 5.5/0.5/6, 6.5/0.5/7, 7.5/0.5/8,
    7.5/1.5/9, 6.5/1.5/10, 5.5/1.5/11, 4.5/1.5/12, 3.5/1.5/13, 2.5/1.5/14, 1.5/1.5/15, 0.5/1.5/16,
    0.5/2.5/17, 1.5/2.5/18, 2.5/2.5/19, 3.5/2.5/20, 4.5/2.5/21, 5.5/2.5/22, 6.5/2.5/23, 7.5/2.5/24,
    7.5/3.5/25, 6.5/3.5/26, 5.5/3.5/27, 4.5/3.5/28, 3.5/3.5/29, 2.5/3.5/30, 1.5/3.5/31, 0.5/3.5/32,
    0.5/4.5/33, 1.5/4.5/34, 2.5/4.5/35, 3.5/4.5/36, 4.5/4.5/37, 5.5/4.5/38, 6.5/4.5/39, 7.5/4.5/40,
    7.5/5.5/41, 6.5/5.5/42, 5.5/5.5/43, 4.5/5.5/44, 3.5/5.5/45, 2.5/5.5/46, 1.5/5.5/47, 0.5/5.5/48,
    0.5/6.5/49, 1.5/6.5/50, 2.5/6.5/51, 3.5/6.5/52, 4.5/6.5/53, 5.5/6.5/54, 6.5/6.5/55, 7.5/6.5/56,
    7.5/7.5/57, 6.5/7.5/58, 5.5/7.5/59, 4.5/7.5/60, 3.5/7.5/61, 2.5/7.5/62, 1.5/7.5/63, 0.5/7.5/64
  }{
    \node[circle,fill=white,draw=black,inner sep=1pt] at (\x,\y) {\tiny \n};
  }

  \node at (4,-0.5) {(b) snake Curve};
\end{scope}

\end{tikzpicture}
\caption{Comparison of (a) Hilbert curve and (b) snake curve mappings for an $8 \times 8$ lattice. Numbers indicate the site ordering in the 1D chain.}
\label{fig:mappings_8x8}
\end{figure}

 We compare the Hilbert and snake mappings for an $8\times 8$ lattice in Figure~\ref{fig:mappings_8x8}. The recursive construction of the Hilbert curve ensures that locality is better preserved across scales: sites that are close in 2D remain close in the 1D ordering, except at a small set of quadrant boundaries~\cite{niedermeier1996}. This reduction in effective interaction range directly improves the efficiency of tensor-network simulations by lowering entanglement growth and reducing the required MPO bond dimension. As our results will show, these advantages become especially pronounced at larger system sizes and in parameter regimes where correlations are stronger.

% Formally, the Hilbert curve bounds the 1D separation of 2D nearest neighbors by $3\sqrt{2N+1}$, where $N=n^2$ is the number of lattice sites. By comparison, the snake mapping produces worst-case separations of order $\mathcal{O}(n)$ \textcolor{red}{Non dovrebbe essere $\mathcal{O}(N)$? Perché anche Hilbert, che è $3\sqrt{2N+1}=3\sqrt{2n^2+1}$, è $\mathcal{O}(n)$}. 

% Requires only: \usepackage{tikz}

\subsection{Tensor Network Implementation}

We employ the MPS formalism to represent the quantum many-body wavefunction of the mapped one-dimensional fermionic system. For an $n \times n$ lattice with $N = n^2$ sites, we define a bijective mapping $\mathcal{M}: \mathbb{Z}^2 \rightarrow \mathbb{Z}$ such that each 2D coordinate $(i,j)$ is mapped to a unique position $\mu$ in the 1D chain:

\begin{equation}
\mathcal{M}: (i,j) \mapsto \mu, \quad \text{where} \quad i,j \in \{0,1,...,n-1\}, \quad \mu \in \{0,1,...,N-1\}
\label{eq:mapping}
\end{equation}

For space-filling curve mapping $\mathcal{M}$, each site $\mu$ in the effective 1D MPS chain corresponds to a site $(i,j)$ of the original 2D lattice, retaining the full fermionic Hilbert space with spin-up and spin-down electrons.

The MPS ansatz~\cite{white1992,schollwock2011,verstraete2008} expresses the wave function as
\begin{equation}\label{EQMPS}
|\Psi\rangle = \sum_{\{s_\mu\}} A^{s_1} A^{s_2} \cdots A^{s_N} \, |s_1,s_2,\ldots,s_N\rangle ,
\end{equation}
where $s_\mu$ denotes the local basis at site $\mu$. For the Hubbard model this basis consists of four states,
\begin{align}
s_\mu \in \{\,|0\rangle_\mu,\;|\uparrow\rangle_\mu,\;|\downarrow\rangle_\mu,\;|\uparrow\downarrow\rangle_\mu \,\},
\end{align}
representing the empty site, single occupancies with spin up $\sigma =\, \uparrow$ or down $\sigma =\,  \downarrow$, and double occupancy. Fermionic creation and annihilation operators $c_{\mu\sigma}^{\dagger},c_{\mu\sigma}$ act on these states according to the usual anticommutation relations, which we enforce via parity-aware tensor legs and fermionic swap gates~\cite{corboz2009fermionic}. Throughout this work, we enforce the conservation of Abelian charges $(N_\uparrow,N_\downarrow)$ in MPS tensors to reduce cost and improve stability.
Each tensor $A^{s_\mu} = A^{s_\mu}_{\alpha_{\mu-1},\alpha_\mu}$ appearing in Eq.~(\ref{EQMPS}) carries one physical index $s_\mu$ and two virtual indices $\alpha_{\mu-1},\alpha_\mu$ of dimension up to the bond dimension $m$. The virtual indices mediate quantum correlations between different sites, while the bond dimension $m$ controls the amount of entanglement that the MPS can faithfully represent. For open boundary conditions, we set $\alpha_0=\alpha_N=1$, so the first and last tensors are effectively matrices rather than rank-3 tensors. Larger values of $m$ systematically improve accuracy at the cost of greater computational effort. The efficiency of the MPS representation is therefore dictated by the entanglement structure of the state, which in turn depends on how well the 2D-to-1D mapping preserves locality.

The Hamiltonian is encoded as a Matrix Product Operator (MPO). The long-range hopping terms induced by the mapping are represented exactly in the MPO. Although this increases the MPO bond dimension relative to a purely nearest-neighbor 1D model, the growth remains tractable for the system sizes studied. All tensor network simulations are implemented using the ITensor library~\cite{itensor}, which natively supports fermionic degrees of freedom and efficient construction of long-range MPOs.

\subsection{Optimization Algorithm}

Our main goal is to obtain the ground state of the Hubbard model in different regimes. To achieve this goal, we employ the two-site Density Matrix Renormalization Group (DMRG) algorithm~\cite{white1992,schollwock2011}, augmented with the Hubig subspace expansion strategy for improved robustness \cite{hubig2015strictly}. In DMRG, the many-body wave function is represented as an MPS and optimized variationally by sweeping back and forth across the chain. At each step, a block of two neighboring sites is optimized to minimize the Rayleigh quotient,
\begin{equation}
E = \frac{\langle \Psi | H | \Psi \rangle}{\langle \Psi | \Psi \rangle},
\end{equation}
after which the updated bond is truncated via singular value decomposition (SVD) to enforce the target bond dimension $m$.  

The two-site update allows for dynamic adjustment of the bond dimension and avoids some of the local minima that can hinder single-site DMRG. The subspace expansion further stabilizes optimization by temporarily expanding the variational space with carefully chosen vectors, accelerating convergence without significantly increasing cost~\cite{hubig2015strictly}.

We monitor convergence using two criteria: the truncation error (discarded weight) $\epsilon_{\text{trunc}}$ and the energy difference between successive sweeps. A calculation at fixed $m$ is considered converged once
$
\epsilon_{\text{trunc}} < 10^{-7},
$
in which case the energy typically stabilizes to within numerical precision. If this condition is met before the maximum number of sweeps, the calculation is terminated early to save computational resources. This adaptive strategy ensures efficient use of the bond dimension while systematically tightening the variational ansatz.

As an illustration, simulations reaching a maximum bond dimension of $m=4000$ for $4\times4$ and $8\times8$ lattices were performed using up to 18 sweeps, with bond dimensions increasing according to the schedule:
\[
m \in \{25, 50, 100, 150, 200, 300, 400, 600, 800, 1200, 1600, 2000, 2000, 3000, 3000, 4000, 4000, 4000\}.
\]
This gradual growth balances accuracy and stability, allowing the algorithm to capture entanglement progressively while avoiding convergence issues that often arise from abrupt bond-dimension jumps.

To validate our findings, we compare selected results against auxiliary-field quantum Monte Carlo benchmarks~\cite{qin2016benchmark} for grid sizes up to \(16 \times 16\), as well as against recent high-accuracy projected entangled pair state (PEPS) calculations~\cite{liu2025peps}, which provide competitive variational energies for comparable system sizes, serving as independent reference points for our tensor network simulations.

\section{\label{sec:results} Results and Discussion}

Our work focuses on the solution of the 2D Hubbard model given by:
\begin{equation}
H = -t \sum_{\langle i,j \rangle, \sigma} \left( c_{i\sigma}^\dagger c_{j\sigma} + \text{h.c.} \right) 
+ U \sum_i n_{i\uparrow} n_{i\downarrow}
\label{eq:H_hubbard}
\end{equation}
where $t$ is the hopping parameter, $U$ represents the on-site Coulomb repulsion, $n_{i\sigma} = c_{i\sigma}^\dagger c_{i\sigma}$ is the number operator for spin $\sigma$ electrons at site $i$, and $c_{i\sigma}^{(\dagger)}$ are fermionic annihilation (creation) operators for spin $\sigma = \uparrow,\downarrow$ at site $i$.

The central idea of our approach is to map the two-dimensional square lattice onto a one-dimensional chain, enabling the use of highly efficient 1D tensor network algorithms. Under this mapping described in Eq \ref{eq:mapping}, the original 2D Hubbard Hamiltonian transforms into a 1D model with both nearest-neighbor and long-range interactions:

% For an $n \times n$ lattice with $N = n^2$ sites, we define a bijective mapping $\mathcal{M}: \mathbb{Z}^2 \rightarrow \mathbb{Z}$ such that each 2D coordinate $(i,j)$ is mapped to a unique position $\mu$ in the 1D chain:

% \begin{equation}
% \mathcal{M}: (i,j) \mapsto \mu, \quad \text{where} \quad i,j \in \{0,1,...,n-1\}, \quad \mu \in \{0,1,...,N-1\}
% \label{eq:mapping}
% \end{equation}

% Under this mapping, the original 2D Hubbard Hamiltonian transforms into a 1D model with both nearest-neighbor and long-range interactions:

\begin{equation}
H_{\mathcal{M}} = -t \sum_{\substack{\mu,\nu \\ \langle\mathcal{M}^{-1}(\mu),\mathcal{M}^{-1}(\nu)\rangle_{2D}}} \sum_{\sigma} \left(c_{\mu\sigma}^{\dagger}c_{\nu\sigma} + \text{h.c.}\right) + U \sum_{\mu} n_{\mu\uparrow}n_{\mu\downarrow}
\end{equation}
where $\langle\cdot,\cdot\rangle_{2D}$ denotes nearest neighbors in the original 2D lattice. The efficiency of tensor network simulations critically depends on the decay of interactions in the mapped 1D system, making the choice of mapping crucial.

\subsection{Open Boundary Conditions (OBC) at Half-Filling (\(\langle n_{\text{occ}}\rangle=1\))}

We begin our analysis under OBC and in the half-filling regime, where the average electron density per site is
\[
n_{\text{occ}} = \frac{1}{N} \Big\langle \sum_{i,\sigma} n_{i\sigma} \Big\rangle = 1,
\]
with \(N=n^2\) the total number of lattice sites. In this setup, the hopping term is restricted to nearest-neighbor couplings within the finite $n \times n$ lattice, with no wrap-around connections in either directions. Consequently, edge and corner sites have fewer nearest neighbors compared to bulk sites, making OBC a natural setting for assessing how well different mappings preserve locality without the additional complications of periodic wrap-around terms.

To quantify the effect of the mapping, we define the relative energy difference between snake and Hilbert orderings as follows:
\begin{equation}
\Delta = \Delta E/|E_{\mathrm{Hilbert}}| = \frac{E_{\mathrm{snake}} - E_{\mathrm{Hilbert}}}{|E_{\mathrm{Hilbert}}|} \times 100\% ,
\label{eq:Delta}
\end{equation}
where $E_{\mathrm{snake}}$ and $E_{\mathrm{Hilbert}}$ denote the ground-state energies per site obtained with the snake and the Hilbert mapping.

\begin{figure}[H]
    \centering
    \includegraphics[width=0.6\textwidth]{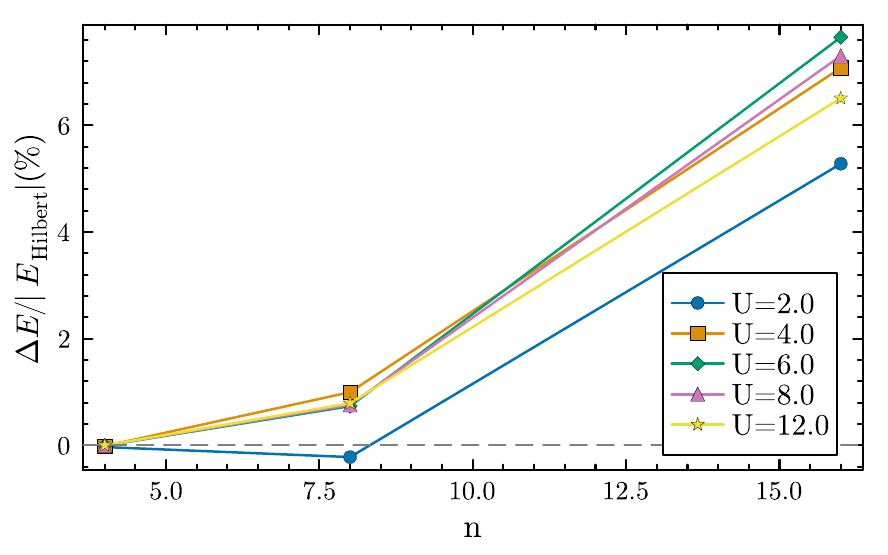}
    % Placeholder for energy difference vs system size
    \caption{Relative energy difference 
$\Delta E/|E_{\mathrm{Hilbert}}|$ between snake and Hilbert mappings for the half-filled Hubbard model under OBC at bond dimension $m=1000$. Results are shown for $4\times4$, $8\times8$, and $16\times16$ lattices as a function of $U$, where $t=1$.}

    \label{fig:deleta_e_different_u_obc}
\end{figure}

We show $\Delta E/|E_{\mathrm{Hilbert}}|$ for system sizes $4\times4$, $8\times8$, and $16\times16$ across the interaction range $U/t\in [2,12]$ in Figure~\ref{fig:deleta_e_different_u_obc}. In all cases, the Hilbert mapping yields consistently lower energies than the snake mapping, confirming its superior efficiency in capturing the ground state at fixed bond dimension. The advantage peaks in the intermediate coupling regime $4 \lesssim U/t \lesssim 8$, where the correlation effects are strongest and the numerical simulations are the most challenging. In this region, the Hilbert curve achieves around $\sim 7\%$ improvement in energy accuracy, underscoring its value for studying the physically relevant crossover from weak to strong coupling.

We focus on the representative intermediate coupling regime around $U/t \approx 6$, which has been extensively studied in the 2D Hubbard literature. This parameter region is of particular interest because it lies near the onset of pseudogap behavior and the regime of strongest superconducting tendencies identified in cluster DMFT/DCA analyses~\cite{gull2012energetics,gull2013scpg}. It has also been investigated in recent finite-temperature Monte Carlo studies that map out pseudogap phenomenology~\cite{jiang2022mc_pg}, and even serves as a calibrated operating point in cold-atom realizations of the Fermi–Hubbard model~\cite{chiu2018coldatoms}, making it a broadly adopted benchmark across numerical and experimental platforms.  

\begin{figure}[htb]
    \centering

    \begin{subfigure}[t]{0.49\textwidth}
        \centering
        \includegraphics[width=\linewidth]{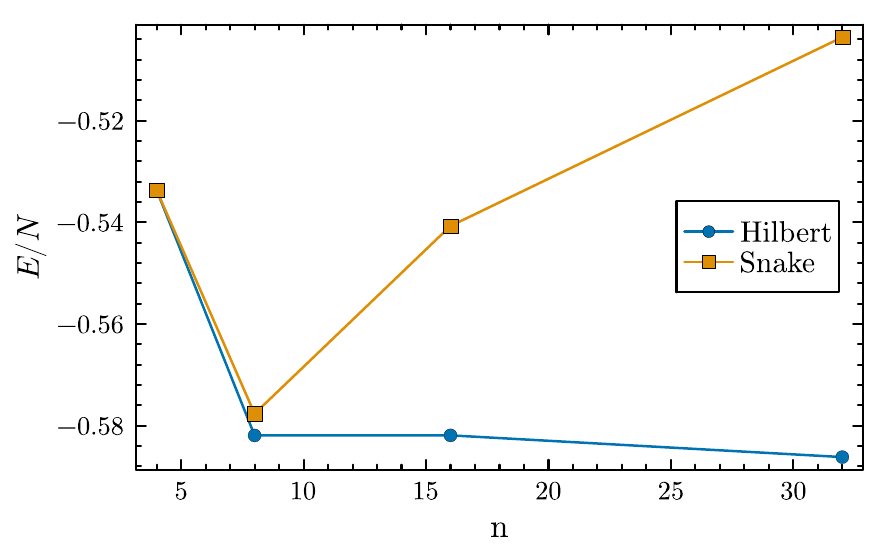}
        \caption{}\label{fig:energy_over_n_a}
    \end{subfigure}\hfill
    \begin{subfigure}[t]{0.49\textwidth}
        \centering
        \includegraphics[width=\linewidth]{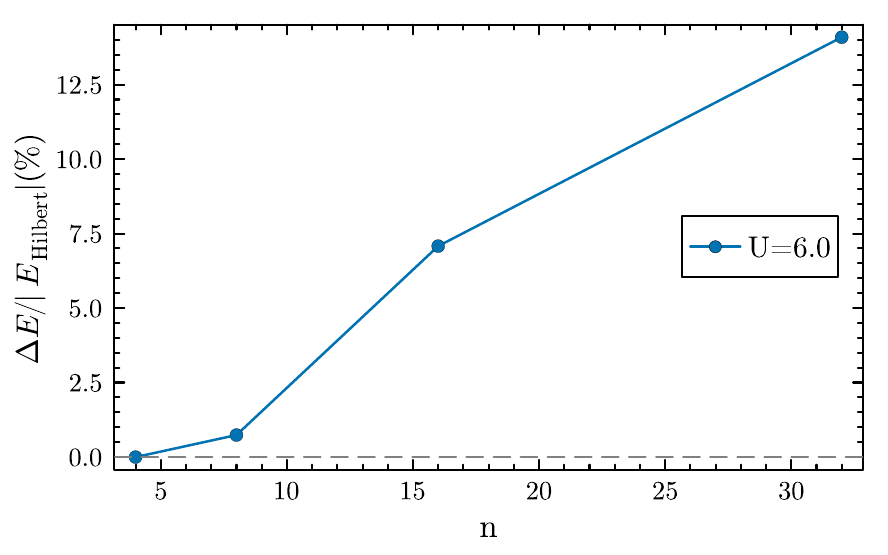}
        \caption{}\label{fig:energy_over_n_b}
    \end{subfigure}

    \caption{(a) Energy per site for Hilbert (blue) and snake (yellow) as a function of system size for the half-filled Hubbard model with $U = 6$ and $ t = 1$. (b) The increasing trend of the gap between the two demonstrates the growing advantage of the Hilbert mapping for larger systems. At fixed bond dimension $m=1000$, the energy-per-site difference grows from $0$ for $4\times4$ to $0.085$ for $32\times32$.}
    \label{fig:energy_over_n_obc}
\end{figure}

To clearly assess the advantage of Hilbert mapping over snake mapping, we concentrate on $U/t=6$, where our simulations show the largest systematic gap between the two approaches. Figure~\ref{fig:energy_over_n_a} presents the ground-state energy per site as a function of linear system size $n$ for both mappings at fixed bond dimension $m=1000$ under OBC. At this modest $m$, the snake mapping yields noticeably higher energies as $n$ increases, reflecting the longer effective distances between vertical neighbors at row boundaries, which enhance entanglement and MPO complexity. In contrast, the Hilbert curve preserves the locality more effectively and consistently produces lower energies for $n=16$ and $n=32$, indicating a more accurate approximation to the ground state within the same computational budget.

The growing advantage of the Hilbert mapping with system size is quantified in Figure~\ref{fig:energy_over_n_b}, which plots $\Delta E/|E_{\mathrm{Hilbert}}|$ across $4\times4$ to $32\times32$ systems at $U/t=6$ and $m=1000$, the relative gap increases monotonically, highlighting that (i) Hilbert’s improved locality translates directly into a lower entanglement budget for MPS, and (ii) the benefit compounds with size precisely in the regime where tensor network costs typically explode. Together with the broader literature identifying $U/t\approx 5.8\!-\!6.4$ as the window where pseudogap features and superconducting correlations are strongest~\cite{gull2012energetics,gull2013scpg,jiang2022mc_pg}, these results underscore that space-filling-curve choices are most consequential exactly where the physics is most delicate.

\begin{figure}[htb]
    \centering

    \begin{subfigure}[t]{0.49\textwidth}
        \centering
        \includegraphics[width=\linewidth]{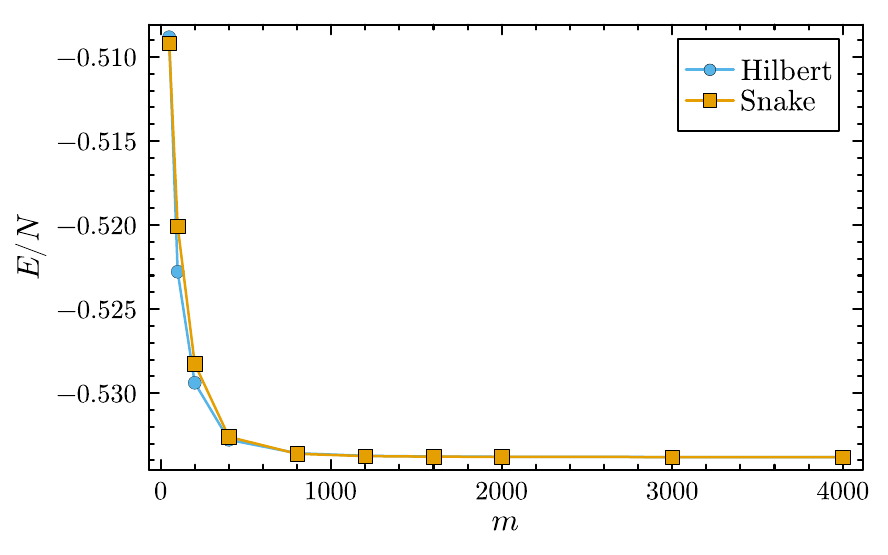}
        \caption{$4 \times 4$}\label{fig:energies_half_filling_a}
    \end{subfigure}\hfill
    \begin{subfigure}[t]{0.49\textwidth}
        \centering
        \includegraphics[width=\linewidth]{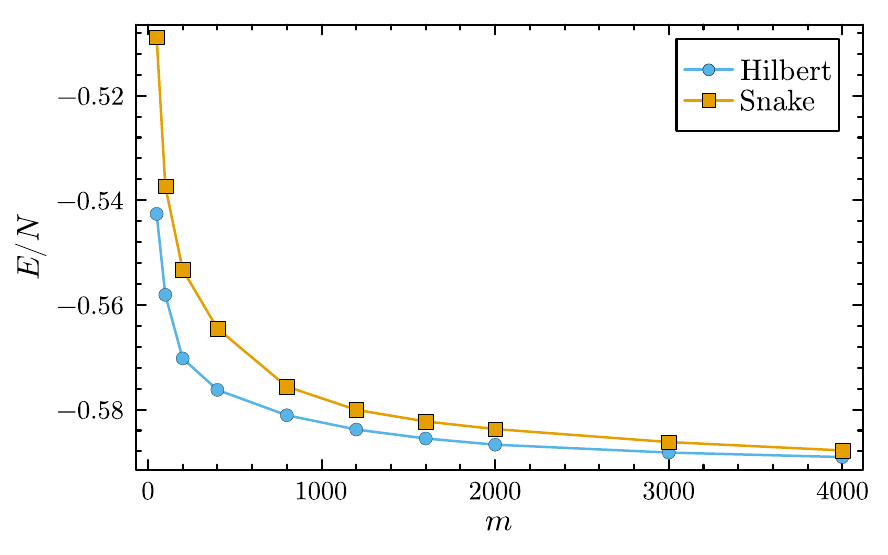}
        \caption{$8 \times 8$}\label{fig:subim2}
    \end{subfigure}

    \medskip

    \begin{subfigure}[t]{0.49\textwidth}
        \centering
        \includegraphics[width=\linewidth]{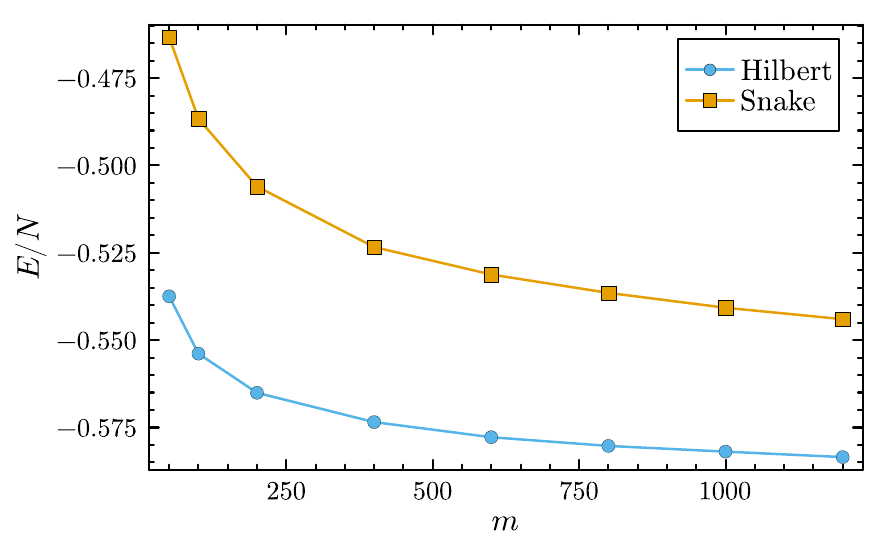}
        \caption{$16 \times 16$}
    \end{subfigure}

    % \begin{subfigure}[t]{0.49\textwidth}
    %     \centering
    %     \includegraphics[width=\linewidth]{32_obc.png}
    %     \caption{$32 \times 32$}\label{fig:subim3}
    % \end{subfigure}

    \caption{Ground state energy per site for the half-filled Hubbard model (OBC) with $U = 6$ and $ t = 1$ as a function of bond dimension for different lattice sizes: (a) $4 \times 4$, (b) $8 \times 8$, and (c) $16 \times 16$. Results are shown for both Hilbert (blue) and snake (yellow) mappings.}
    \label{fig:energies_half_filling}
\end{figure}

In Figure~\ref{fig:energies_half_filling}, we present the ground state energy per site as a function of the bond dimension for $U/t=6$ under OBC on the $4\times4$, $8\times8$, and $16\times16$ lattices. In all cases, the energy decreases monotonically with an increase in the bond dimension, as expected for a variational approach. Across all system sizes, at a fixed bond dimension, the Hilbert mapping yields lower energies than the snake mapping, confirming its more locality-preserving nature. 

\subsection{Periodic Boundary Conditions (PBC) at Half-Filling (\(\langle n_{\text{occ}}\rangle=1\))}

In this section, we propose an analysis of PBC in the half-filling regime. In this setup, the hopping term connects not only nearest-neighbor sites within the bulk but also sites across the lattice edges, effectively wrapping the square lattice into a torus topology. As a result, every site has exactly four neighbors, eliminating edge effects that are present under OBC. The on-site interaction term $U\sum_i n_{i\uparrow}n_{i\downarrow}$ remains strictly local and is unaffected by the boundary choice.

From a numerical point of view, PBC introduces additional challenges for tensor network simulations. While the Hamiltonian remains local on the 2D torus, the 2D-to-1D mapping generates couplings between sites that are far apart in the one-dimensional ordering, particularly at the wrap-around boundaries. These long-range terms lead to higher entanglement across the MPS bonds and thus require significantly larger bond dimensions to achieve convergence.

\begin{figure}[htb]
    \centering

    \begin{subfigure}[t]{0.49\textwidth}
        \centering
        \includegraphics[width=\linewidth]{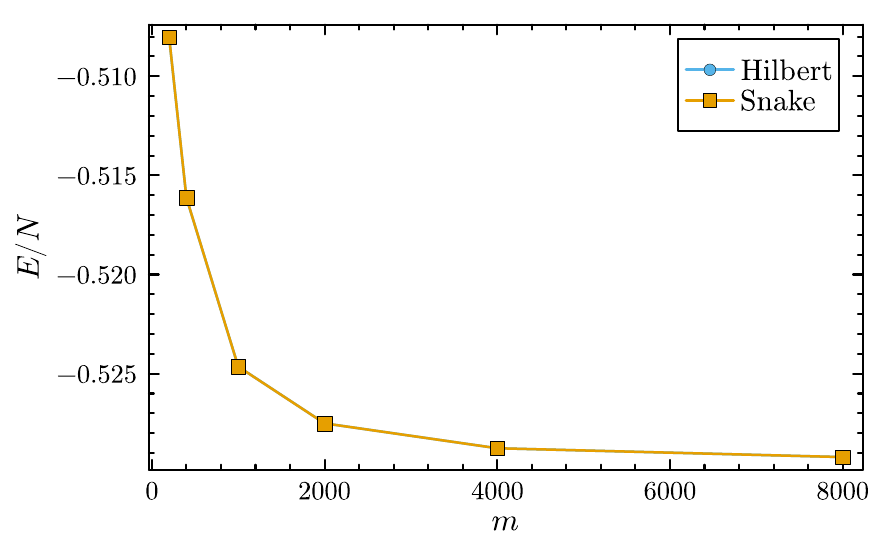}
        \caption{$4 \times 4$}\label{fig:subim1}
    \end{subfigure}\hfill
    \begin{subfigure}[t]{0.49\textwidth}
        \centering
        \includegraphics[width=\linewidth]{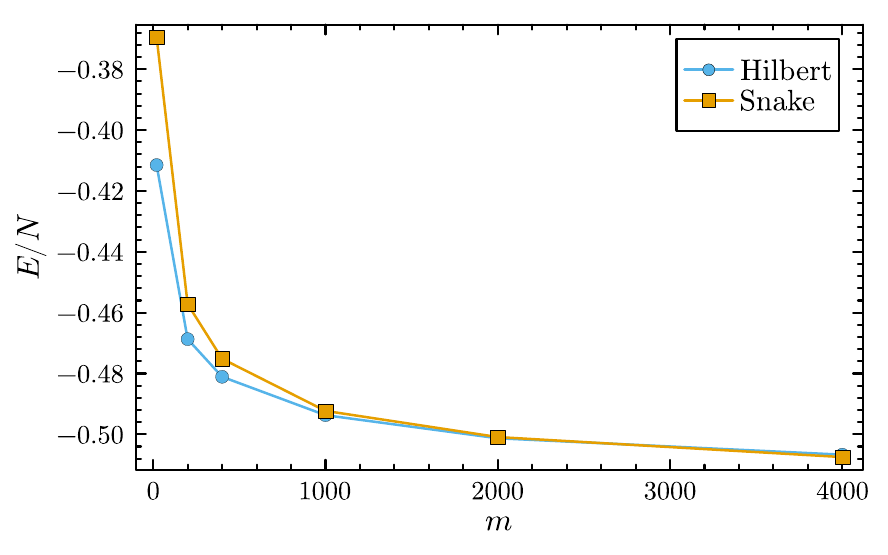}
        \caption{$8 \times 8$}\label{fig:subim2}
    \end{subfigure}

    \medskip

    \begin{subfigure}[t]{0.49\textwidth}
        \centering
        \includegraphics[width=\linewidth]{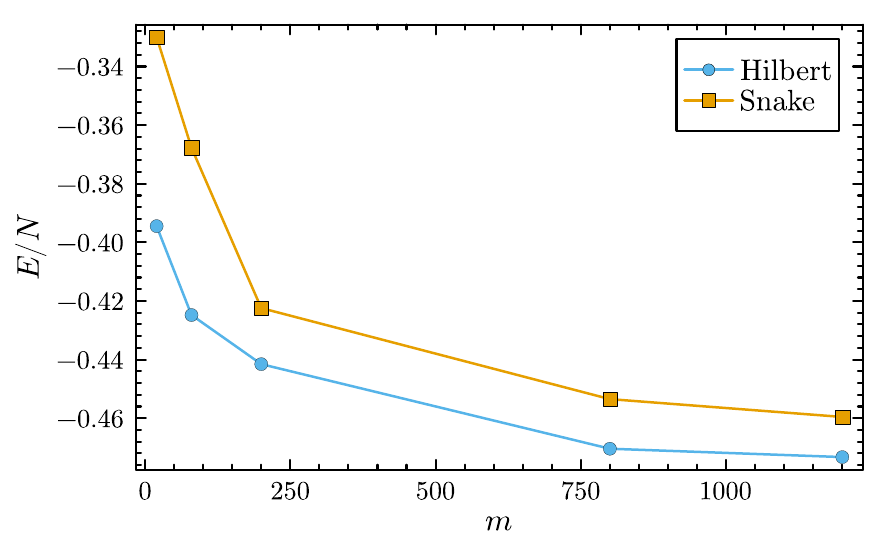}
        \caption{$16 \times 16$}
    \end{subfigure}

    % \begin{subfigure}[t]{0.49\textwidth}
    %     \centering
    %     \includegraphics[width=\linewidth]{32_pbc.png}
    %     \caption{$32 \times 32$}\label{fig:subim3}
    % \end{subfigure}
    
    \caption{Ground-state energy per site for the half-filled Hubbard model with $U/t=8$ under PBC. Results are shown as a function of bond dimension $m$ for (a) $4\times4$, (b) $8\times8$, and (c) $16\times16$ lattices. The Hilbert mapping (blue) consistently outperforms the snake mapping (yellow), with the performance gap widening as the system size increases.}

    % \textcolor{red}{The horizontal dashed lines indicate the benchmark values from ....~\cite{leblanc2015}.} }
    
    \label{fig:e_over_m_pbc}
\end{figure}

In Figure~\ref{fig:e_over_m_pbc}, we show the energy of the ground state per site as a function of the bond dimension for $U/t=8$ on the $4\times4$, $8\times8$, and $16\times16$ lattices. Crucially, for any fixed bond dimension the Hilbert mapping consistently achieves lower energies than the snake mapping, demonstrating its more efficient encoding of the ground state. This advantage becomes more pronounced as the system size grows, reflecting the Hilbert curve’s superior ability to preserve locality even in the presence of PBC-induced wrap-around couplings.

\begin{figure}[htb]
    \centering

    \begin{subfigure}[t]{0.49\textwidth}
        \centering
        \includegraphics[width=\linewidth]{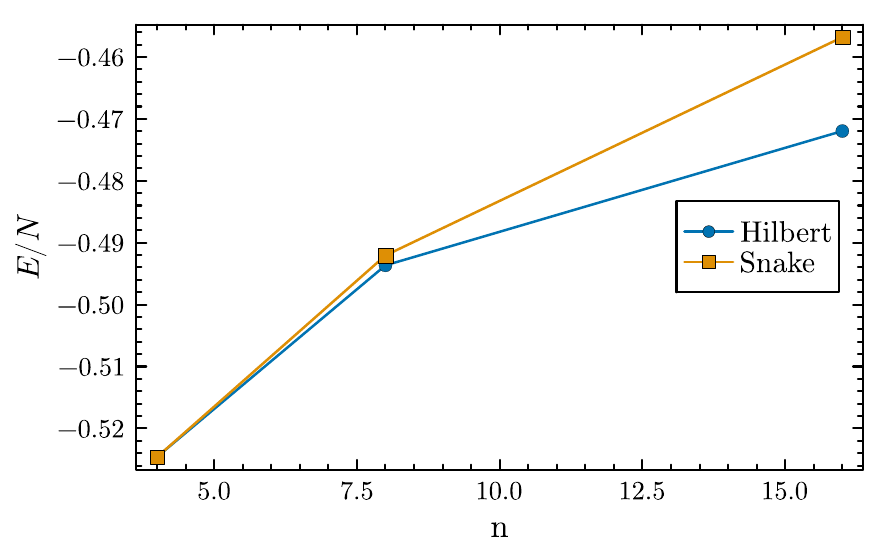}
        \caption{}\label{fig:energy_difference_pbc_a}
    \end{subfigure}\hfill
    \begin{subfigure}[t]{0.49\textwidth}
        \centering
        \includegraphics[width=\linewidth]{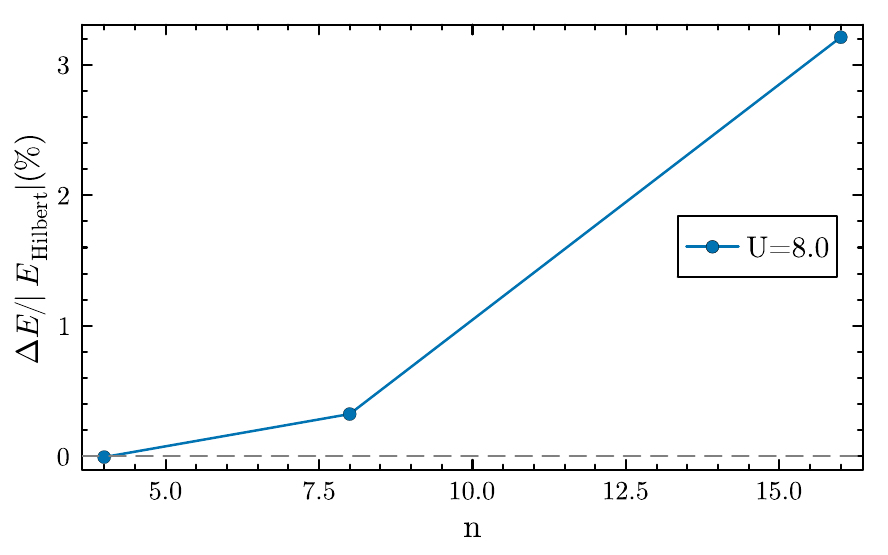}
        \caption{}\label{fig:energy_difference_pbc_b}
    \end{subfigure}

    \caption{(a) Energy per site for Hilbert (blue) and snake (yellow) as a function of system size for the half-filled Hubbard model (PBC) with $U/t = 8$. (b) The increasing trend of the gap between the two demonstrates the growing advantage of the Hilbert mapping for larger systems. At fixed bond dimension $m=1000$, the energy-per-site difference grows from $0$ for $4\times4$ to $0.015$ for $16\times16$.}
    \label{fig:energy_difference_pbc}
\end{figure}

Figure~\ref{fig:energy_difference_pbc_a} compares the ground-state energy per site obtained with Hilbert and snake mappings for increasing lattice sizes. In contrast to the OBC case, both mappings show a systematic drift of the estimated energies with system size yet we consistently find that the Hilbert mapping achieves lower variational energies than the snake mapping at every size studied, and that the gap between the two grows with increasing system size.

To make the comparison more quantitative, Figure~\ref{fig:energy_difference_pbc_b} plots the relative energy difference 
$\Delta E/|E_{\mathrm{Hilbert}}|$ as a function of system size. The positive and growing trend demonstrates that the Hilbert mapping consistently achieves lower energies and that its relative advantage increases with lattice size. This scaling behavior highlights that the Hilbert curve provides a more locality-preserving representation of the 2D Hubbard model under PBC, delaying the onset of finite-bond-dimension breakdown compared to the conventional snake ordering.

To contextualize our results, we note that LeBlanc et al. ~\cite{leblanc2015} provide a widely used benchmark suite for the 2D Hubbard model, which includes DMRG data on cylindrical boundary conditions, namely periodic in one direction and open in the other. Because those DMRG results are obtained on nonsymmetric geometries, snake mapping can gain a topological advantage that stems from the cylindrical shape, making a direct comparison with our strictly periodic or open boundary conditions misleading. We therefore do not attempt to compare our PBC/OBC energies against those cylindrical DMRG data.

Qin et al.~\cite{qin2016benchmark} provided accurate auxiliary-field quantum Monte Carlo benchmarks for finite clusters up to \(16 \times 16\) with full periodic boundary conditions. Their results at \(U/t=8\) and half-filling offer a direct and reliable reference for our PBC calculations. We find that our \(4 \times 4\) values nearly coincide with the benchmark, while for \(8 \times 8\) both mappings deviate by about \(4\%\). However, as the system size increases, the Hilbert mapping deviates more slowly from the QMC reference than the snake mapping. For \(16 \times 16\), the snake deviates by roughly \(13\%\), whereas the Hilbert mapping shows a smaller deviation of around \(10\%\), indicating that its locality-preserving structure helps mitigate deterioration at larger sizes.

We emphasize that, even for 2D PBC, we employ an MPS with open virtual boundaries and encode wrap-around couplings explicitly in the MPO, a standard practice that avoids the known instabilities of PBC-MPS while preserving the torus Hamiltonian exactly.

\subsection{Doped System}

\begin{figure}%[H]
    \centering
    \includegraphics[width=0.5\textwidth]{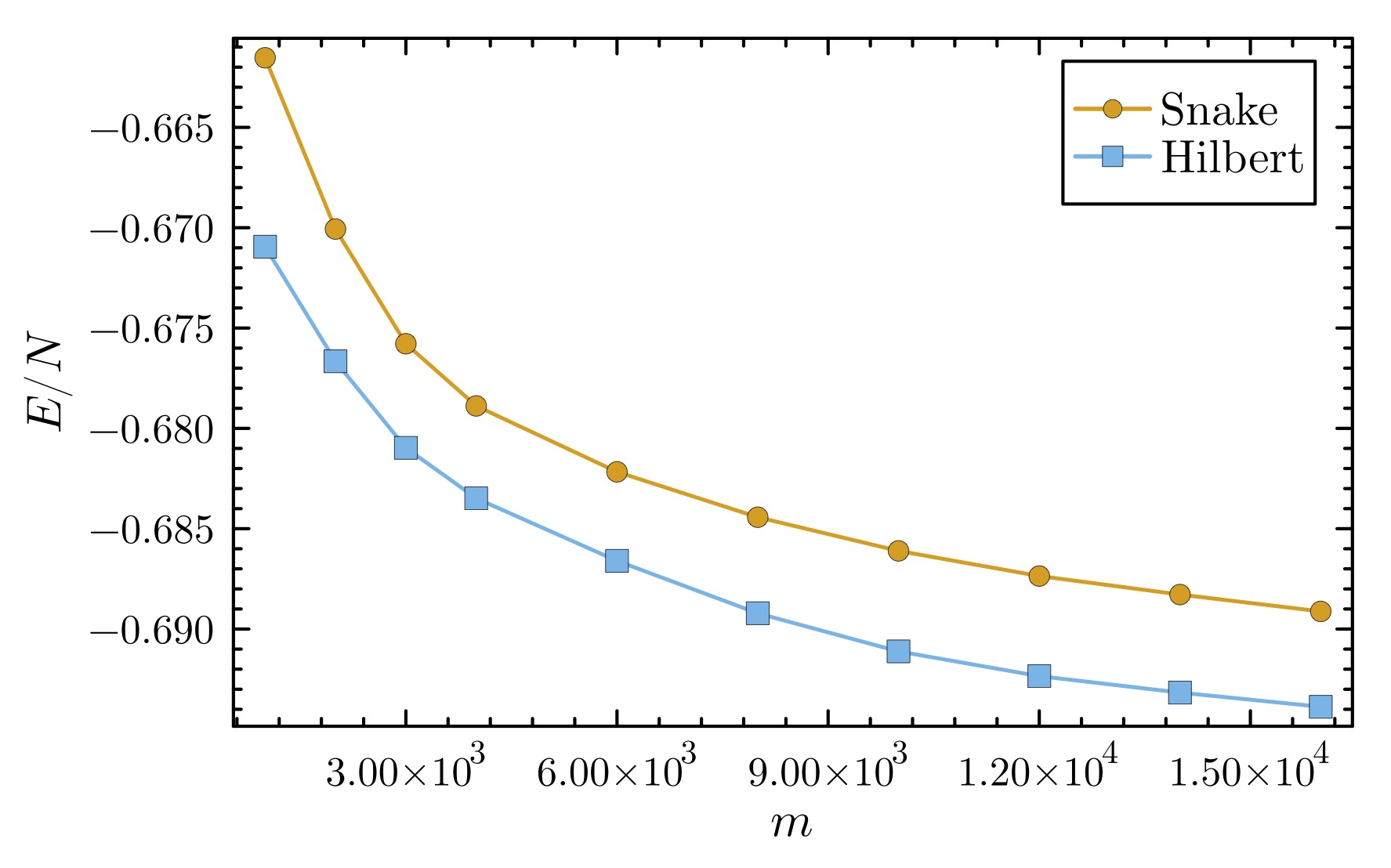}
    \caption{Ground-state energy per site for the doped Hubbard model (\(n = 0.875\), \(U/t = 8\)) on an \(8 \times 8\) lattice under OBC. The Hilbert mapping (blue) consistently achieves lower energies than the snake mapping (yellow) at comparable bond dimensions, demonstrating its superior efficiency and accuracy in the doped regime. For \(m = 16000\), the energies per site are \(E/N = -0.6891\) (snake) and \(E/N = -0.6939\) (Hilbert).}
    \label{fig:doped}
\end{figure}

Away from half-filling, the Hubbard model exhibits even richer physics, including stripe order and $d$-wave superconductivity. The stripe order refers to the spontaneous formation of unidirectional charge and spin density modulations, while $d$-wave superconductivity corresponds to Cooper pairing with a gap function that changes sign under $90^\circ$ rotations. Both phenomena have been widely discussed as possible consequences of strong correlations in the doped Hubbard model~\cite{kivelson2003electron,stripe,leblanc2015}. We study the system at 12.5\% hole doping (density $n = 0.875$) with $U/t = 8$ and under OBC. At this doping level, QMC methods suffer from the sign problem~\cite{leblanc2015,mondaini2022,qin2022science}, making tensor network approaches particularly valuable. Figure~\ref{fig:doped} shows the energy per site of Hilbert and snake mapping in this doped regime.

In particular, for an \(8 \times 8\) lattice with eight holes (12.5\% doping), the Hilbert mapping at bond dimension \(m = 16000\) yields a ground state energy approximately \(0.7\%\) lower than that obtained with the snake mapping at the same \(m\). Moreover, our results lie within \(1\%\) of the state-of-the-art finite PEPS benchmarks reported by Liu \emph{et al.}~\cite{liu2025peps}. Figure~\ref{fig:doped} highlights the superior performance of the Hilbert mapping even at high bond dimensions in doped systems.

\subsection{Tree Tensor Networks (TTN) from Hilbert-Inspired Topologies}

Although Hilbert mapping significantly improves MPS simulations by preserving locality, it has an inherent limitation: sites at the boundaries between quadrants can still experience long-range interactions in the 1D chain. Specifically, when the Hilbert curve transitions between the main quadrants of the 2D lattice, neighboring sites in 2D can be separated by distances of order $\mathcal{O}(\sqrt{N})$ in the 1D ordering. This motivates us to explore alternative tensor network structures that can better accommodate these boundary discontinuities.

We propose a class of tree tensor networks (TTNs) whose connectivity is inspired by the hierarchical structure of the Hilbert curve. TTNs are loop-free tensor networks in which internal tensors connect disjoint blocks of sites hierarchically~\cite{shi2006ttn,orus2014intro,nakatani2013ttns}. Unlike a strict 1D chain (MPS), these TTNs introduce branching points at strategic locations corresponding to quadrant boundaries in the Hilbert mapping. This approach leverages the self-similar, fractal nature of the Hilbert curve to construct a network topology that more naturally reflects the 2D connectivity.

The construction begins with the standard Hilbert mapping for a $n \times n$ lattice (where $n = 2^k$). We identify critical junction points where the curve transitions between major quadrants - these typically correspond to sites with the longest-range interactions in the 1D chain. At these junctions, we modify the tensor network structure by introducing three-leg tensors (branching nodes) that connect multiple quadrants, removing certain long-range connections in favor of shorter tree branches while maintaining overall connectivity such that the resulting structure remains a tree without loops. This produces a Hilbert-inspired TTN that preserves the recursive structure of the original mapping while shortening several critical graph distances responsible for the longest-range terms.

\begin{figure}[htb]
\centering
\begin{tikzpicture}[scale=0.7]
% TTN-A (left)
\begin{scope}[xshift=0cm]
\draw[step=1cm,gray,very thin] (0,0) grid (4,4);

% Main path with central branch
\draw[blue,very thick,->] 
(3.5,0.5) -- (2.5,0.5) -- (2.5,1.5) -- (3.5,1.5) --
(3.5,2.5) -- (3.5,3.5) -- (2.5,3.5) -- (2.5,2.5) --
(1.5,1.5) --
(0.5,1.5) -- (0.5,2.5) -- (0.5,3.5) -- (1.5,3.5) -- (1.5,2.5);

% Branch to bottom
\draw[blue,very thick]
(1.5,1.5) -- (1.5,0.5) -- (0.5,0.5);

\node at (2,-0.7) {\small (a) TTN-A};

% Node labels
\foreach \x/\y/\n in {3.5/0.5/16, 2.5/0.5/15, 2.5/1.5/14, 3.5/1.5/13,
                      3.5/2.5/12, 3.5/3.5/11, 2.5/3.5/10, 2.5/2.5/9,
                      1.5/2.5/8, 1.5/3.5/7, 0.5/3.5/6, 0.5/2.5/5,
                      0.5/1.5/4, 1.5/1.5/3, 1.5/0.5/2, 0.5/0.5/1} {
    \node[circle,fill=white,draw=black,inner sep=1pt] at (\x,\y) {\tiny \n};
}
% Highlight branching node
\node[circle,fill=yellow!50,draw=black,inner sep=1pt] at (1.5,1.5) {\tiny 3};
\end{scope}

% TTN-B (right)
\begin{scope}[xshift=6cm]
\draw[step=1cm,gray,very thin] (0,0) grid (4,4);

% Top/right run ends at 9
\draw[blue,very thick,->] 
(3.5,0.5) -- (2.5,0.5) -- (2.5,1.5) -- (3.5,1.5) --
(3.5,2.5) -- (3.5,3.5) -- (2.5,3.5) -- (2.5,2.5);

% Left column chain
\draw[blue,very thick]
(1.5,1.5) -- (0.5,1.5) -- (0.5,2.5) -- (0.5,3.5) -- (1.5,3.5) -- (1.5,2.5);

% Branch to bottom
\draw[blue,very thick]
(1.5,1.5) -- (1.5,0.5) -- (0.5,0.5);

% Shortcut: 3 -> 14
\draw[blue,very thick]
(1.5,1.5) -- (2.5,1.5);

\node at (2,-0.7) {\small (b) TTN-B};

% Node labels
\foreach \x/\y/\n in {3.5/0.5/16, 2.5/0.5/15, 2.5/1.5/14, 3.5/1.5/13,
                      3.5/2.5/12, 3.5/3.5/11, 2.5/3.5/10, 2.5/2.5/9,
                      1.5/2.5/8, 1.5/3.5/7, 0.5/3.5/6, 0.5/2.5/5,
                      0.5/1.5/4, 1.5/1.5/3, 1.5/0.5/2, 0.5/0.5/1} {
    \node[circle,fill=white,draw=black,inner sep=1pt] at (\x,\y) {\tiny \n};
}
% Highlight branching nodes
\node[circle,fill=yellow!50,draw=black,inner sep=1pt] at (1.5,1.5) {\tiny 3};
\node[circle,fill=yellow!50,draw=black,inner sep=1pt] at (2.5,1.5) {\tiny 14};
\end{scope}
\end{tikzpicture}
\caption{Tree Tensor Network topologies inspired by the Hilbert curve for a $4 \times 4$ lattice. Blue edges indicate TTN connections; yellow nodes mark branching nodes with 3 legs. (a) TTN-A creates a central branch connecting upper and lower subtiles. (b) TTN-B introduces branches at sites 3 and 14, with a shortcut connection between them. Both designs generalize recursively to larger $2^k \times 2^k$ lattices.}
\label{fig:ttn_h}
\end{figure}
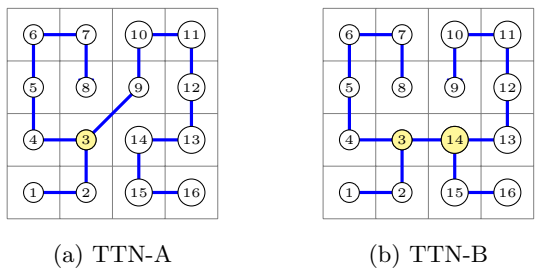

Figure~\ref{fig:ttn_h} shows two minimal variants derived from a $4 \times 4$ Hilbert curve. The configuration TTN-A introduces a central branch connecting the upper-right to the lower-left subtile, while TTN-B creates a branch at a lower junction with a shortcut to the right subtile. Both designs generalize straightforwardly to larger $2^k \times 2^k$ lattices by repeating the same local modification at each scale. Branching remains sparse, only at selected junctions, so the network remains close to an MPS while gaining the power of a hierarchical structure and having fewer long-range interactions.

For a node with  $z$ neighbors, the dominant contractions scale as $\mathcal{O}(m^{z+1})$, where $m$ is the dimension of the bond. Our sparse-branch designs maintain mainly $z=2$ nodes (retaining the $\mathcal{O}(m^3)$ cost of MPS) with only a few $z=3$ junctions that incur $\mathcal{O}(m^4)$ scaling. This overhead is compensated for by the reduced maximum interaction range with respect to the Hilbert-MPS, and the more efficient distribution of the entanglement throughout the network. Different branches can also accommodate different bond dimensions, allowing computational resources to be allocated where correlations are strongest.

\begin{figure}[htb]
    \centering

    \begin{subfigure}[t]{0.49\textwidth}
        \centering
        \includegraphics[width=\linewidth]{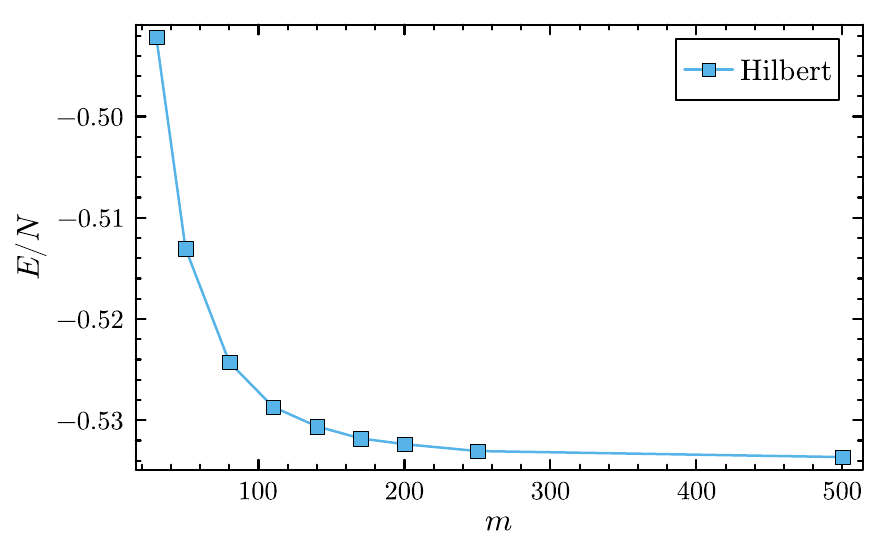}
        \caption{TTN-B results}\label{fig:ttn-b-a}
    \end{subfigure}\hfill
    \begin{subfigure}[t]{0.49\textwidth}
        \centering
        \includegraphics[width=\linewidth]{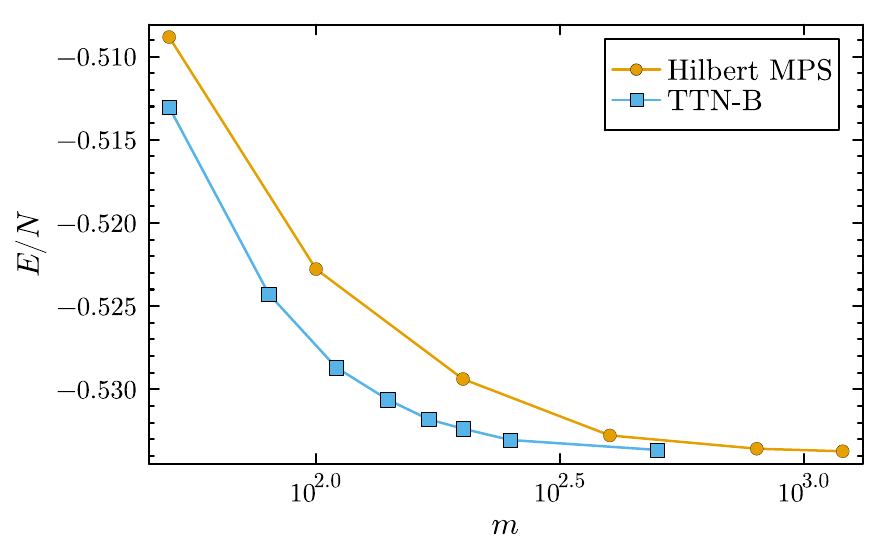}
        \caption{Comparison with Hilbert MPS}\label{fig:ttn-b-b}
    \end{subfigure}

    \caption{Ground-state energy per site for the \(4\times4\) Hubbard model at \(U/t = 6\) under OBC. (a) Convergence of the TTN-B topology with bond dimension. (b) Direct comparison between TTN-B (blue) and Hilbert-mapped MPS (yellow, reproduced from Figure~\ref{fig:energies_half_filling_a}) on a logarithmic scale. The TTN-B network achieves the converged MPS energy at bond dimension \(m \simeq 250\), compared to \(m \simeq 1000\) required by the Hilbert MPS.}
    \label{fig:ttn-b}
\end{figure}

Figure~\ref{fig:ttn-b} shows the ground-state energy per site at \(U/t = 6\) under OBC obtained using the TTN-B network topology. Panel (a) displays the TTN-B convergence, while panel (b) provides a direct comparison with the Hilbert-mapped MPS results from Figure~\ref{fig:energies_half_filling_a} using a logarithmic bond-dimension axis. The TTN-B network reproduces the converged MPS energy at a bond dimension of approximately \(m \simeq 250\), compared to \(m \simeq 1000\) required by the Hilbert MPS. As expected, the TTN achieves convergence at bond dimensions higher than PEPS~\cite{Corboz_2016} but lower than MPS, highlighting its potential as an efficient intermediate tensor network structure for exploring ground states of two-dimensional correlated systems.

Importantly, this evolution from MPS to TTN is uniquely enabled by the self-similar structure of the Hilbert curve. The snake mapping, which lacks this hierarchical organization, does not naturally suggest analogous tree structures. This demonstrates an additional advantage of the Hilbert approach: it provides a systematic pathway for constructing more sophisticated tensor network architectures tailored to the specific connectivity patterns of the 2D-to-1D mapping. Preliminary results indicate that these TTN structures can achieve an accuracy comparable to Hilbert-mapped MPS, particularly when the system has intermediate or strong correlations, where the additional flexibility of the tree structure becomes most beneficial. A comprehensive analysis of these structures, including detailed benchmarking against MPS for various system sizes and parameters, could be the subject of future work.

\section{\label{sec:conclusions} Conclusions}

We have shown that mapping the two-dimensional Hubbard model onto a one-dimensional chain via the Hilbert space-filling curve yields a more efficient tensor-network representation than conventional snake ordering. Across diverse regimes, we tested open and periodic boundary conditions at half-filling and representative doped cases. We showed that the Hilbert mapping achieves systematically lower variational energies at fixed bond dimension, with the advantage growing with linear system size. This reflects the superior locality preservation of the curve: 2D nearest-neighbor couplings translate into shorter-range and more evenly distributed interactions in 1D, reducing entanglement between MPS cuts and the effective complexity of MPO.

The benefit is most pronounced in the physically central intermediate coupling window (e.g. \(U/t\!\approx\!6\)), where correlations are strongest and where many numerical methods are most challenged. Even under periodic boundary conditions where 1D encodings suffer the most from wrap-around terms, the Hilbert mapping delays the finite \(m\) breakdown relative to snake order, yielding lower energies at fixed resources.

However, there are fundamental limits: representing a 2D torus with a 1D ansatz still induces long-range terms and entanglement growth that ultimately requires large \(m\). Although our implementation extends accessible clusters (e.g., from narrow cylinders to full \(L\times L\) lattices in the sizes reported here) and competes favorably with alternative tensor-network baselines at comparable cost, reaching the thermodynamic limit with uniform accuracy will require further algorithmic progress.

Our results suggest several concrete directions. First, Hilbert-inspired tree tensor networks (TTN) can target residual long-range 'quadrant-jump' couplings by introducing sparse branching at the hierarchical junctions of the curve, potentially lowering the maximum cut entanglement at modest overhead. Second, automated path optimization, tiling and refining locality-preserving motifs atop the Hilbert, offers a principled way to close the remaining gap between 2D connectivity and 1D representations. Together, these avenues point to scalable tensor-network simulations of larger, more strongly correlated 2D fermionic systems.

\begin{acknowledgments}
VG acknowledges financial support by MUR (Ministero dell'Universit\`a e della Ricerca) through the PNRR MUR project PE0000023-NQSTI.
\end{acknowledgments}

\bibliographystyle{apsrev4-2} 
\bibliography{bib}

\appendix

\end{document}